\documentclass{article}

\usepackage[preprint]{neurips_2022}

\usepackage{xcolor}         % colors
\definecolor{linkColor}{rgb}{0.18,0.39,0.62}
\usepackage[utf8]{inputenc} % allow utf-8 input
\usepackage[T1]{fontenc}    % use 8-bit T1 fonts
\usepackage[colorlinks=true,linkcolor=linkColor,citecolor=linkColor,filecolor=linkColor,urlcolor=linkColor]{hyperref}       % hyperlinks
\usepackage{url}            % simple URL typesetting
\usepackage{booktabs}       % professional-quality tables
\usepackage{amsfonts}       % blackboard math symbols
\usepackage{nicefrac}       % compact symbols for 1/2, etc.
\usepackage{microtype}      % microtypography
\usepackage[most]{tcolorbox}
\usepackage{enumitem}

\usepackage{amsmath}
\usepackage{amssymb}
\usepackage{mathtools}
\usepackage{amsthm}
\usepackage{graphicx}
\usepackage{subfigure}
\usepackage{xspace}
\usepackage{pifont}
\usepackage{xcolor}
\usepackage{array}

\theoremstyle{plain}

\theoremstyle{definition}

\theoremstyle{remark}

\usepackage[capitalize,noabbrev]{cleveref}
\usepackage{framed}
\usepackage{multirow}
\usepackage{svg}
\usepackage{listings}
\usepackage{wrapfig}

\DeclareMathOperator*{\argmin}{arg\,min}
\usepackage{multirow}
\usepackage{fontawesome5}

\newcommand*{\img}[1]{%
    \raisebox{-.2\baselineskip}{%
        \includegraphics[
        height=\baselineskip,
        width=\baselineskip,
        keepaspectratio,
        ]{#1}%
    }%
}

\newcommand\our{\textsc{BEATs}}

\newcommand{\cmark}{\ding{51}\xspace}%
\newcommand{\xmark}{\ding{55}\xspace}%

\makeatletter

\newcommand*{\@rowstyle}{}

\newcommand*{\rowstyle}[1]{% sets the style of the next row
  \gdef\@rowstyle{#1}%
  \@rowstyle\ignorespaces%
}

\newcolumntype{=}{% resets the row style
  >{\gdef\@rowstyle{}}%
}

\newcolumntype{+}{% adds the current row style to the next column
  >{\@rowstyle}%
}

\makeatother

\title{\our{}\img{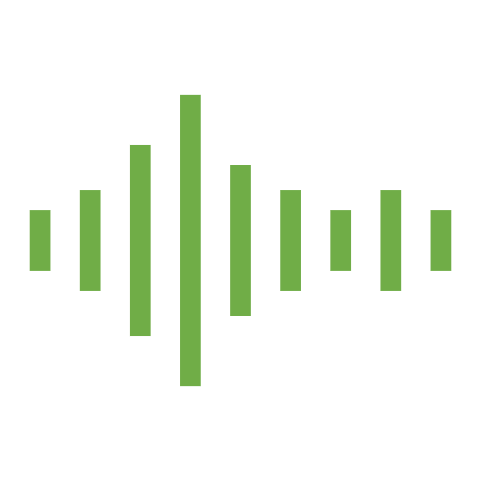}: Audio Pre-Training with Acoustic Tokenizers}

\author{%
  Sanyuan Chen\thanks{This work is done during the internship at MSRA. $\dagger$ Corresponding author } ~\
  Yu Wu$\dagger$~\
  Chengyi Wang~\
  Shujie Liu~\
  Daniel Tompkins~\ \\
  \textbf{Zhuo Chen}~\
  \textbf{Furu Wei} \\
  \{t-schen, yuwu1, t-chewang, shujliu, datompki, zhuc, fuwei\}@microsoft.com  \\
    Microsoft  \\
}

\begin{document}

\maketitle

\begin{abstract}
 The massive growth of self-supervised learning (SSL) has been witnessed in language, vision, speech, and audio domains over the past few years. While discrete label prediction is widely adopted for other modalities, the state-of-the-art audio SSL models still employ reconstruction loss for pre-training.  Compared with reconstruction loss, semantic-rich discrete label prediction encourages the SSL model to abstract the high-level audio semantics and discard the redundant details as in human perception. However, a semantic-rich acoustic tokenizer for general audio pre-training is usually not straightforward to obtain, due to the continuous property of audio and unavailable phoneme sequences like speech. To tackle this challenge, we propose \our{}, an iterative audio pre-training framework to learn \textbf{B}idirectional \textbf{E}ncoder representation from \textbf{A}udio \textbf{T}ransformer\textbf{s}, where an acoustic tokenizer and an audio SSL model are optimized by iterations. In the first iteration, we use random projection as the acoustic tokenizer to train an audio SSL model in a mask and label prediction manner. Then, we train an acoustic tokenizer for the next iteration by distilling the semantic knowledge from the pre-trained or fine-tuned audio SSL model. The iteration is repeated with the hope of mutual promotion of the acoustic tokenizer and audio SSL model. The experimental results demonstrate our acoustic tokenizers can generate discrete labels with rich audio semantics and our audio SSL models achieve state-of-the-art results across various audio classification benchmarks, even outperforming previous models that use more training data and model parameters significantly. Specifically, we set a new state-of-the-art mAP 50.6$\%$ on AudioSet-2M  for audio-only models without using any external data, and 98.1$\%$ accuracy on ESC-50. The code and pre-trained models are available at \url{https://aka.ms/beats}.
\end{abstract}

\section{Introduction}
Recent years have witnessed great success in self-supervised learning (SSL) for speech and audio processing. 
The speech SSL models, such as Wav2vec 2.0 \citep{baevski2020wav2vec}, HuBERT \citep{hsu2021hubert}, BigSSL \citep{zhang2022bigssl}, WavLM \citep{chen2022wavlm}, and data2vec \citep{baevski2022data2vec}, show prominent performance across various speech processing tasks, especially in low-resource scenarios. 
Different from speech, audio typically contains wide variations of environmental events, including human voices, nature sounds, musical beats, etc, which brings great challenges to general audio modeling.
To this end, audio SSL models, such as SS-AST \citep{gong2022ssast} and Audio-MAE \citep{xu2022masked}, are proposed for general audio classification applications, demonstrating that SSL learns robust auditory representations not only for speech but also for non-speech signals. 

Until now, state-of-the-art (SOTA) audio SSL models \citep{xu2022masked,chong2022masked} employ an acoustic feature reconstruction loss as the pre-training objective instead of the discrete label prediction pre-training task as in SSL models of 
speech \citep{hsu2021hubert,chen2022wavlm}, vision \citep{bao2021beit,beitv2,wang2022image} and language \citep{devlin2019bert,liu2019roberta,lan2019albert}. 
However, it was generally believed that the reconstruction loss only accounts for the correctness of low-level time-frequency features  but neglects high-level audio semantic abstraction \citep{ramesh2021zero,bao2021beit}. The discrete label prediction would be a potentially better audio pre-training objective than reconstruction for the following reasons.

Firstly, from the bionics aspect, humans understand audio by extracting and clustering the high-level semantics instead of focusing on the low-level time-frequency details.
For example, humans can easily recognize the sound of any dog barking by capturing and classifying the semantic patterns, even though the dog has never barked in the same scenario with the same tone before. 
By mimicking the semantics extracting and clustering through the discrete label prediction pre-training objective, the audio SSL model is expected to learn the same understanding and generalization skills as humans.

Secondly, from the aspect of modeling efficiency, the reconstruction loss may waste the audio model parameter capacity and pre-training resources on predicting the semantic irrelevant information, which has little benefit to the general audio understanding tasks.
In comparison, the discrete label prediction objective can improve the audio modeling efficiency by providing semantic-rich tokens as the pre-training targets and encouraging the model to discard the redundant details, resulting in a superior audio understanding capability with a lower pre-training recourse cost.

Thirdly, the audio SSL pre-training with the discrete label prediction objective advances the unification of language, vision, speech, and audio pre-training.
Instead of designing the pre-training task for each modality, this unification enables the possibility of building a foundation model across modalities with a single pre-training task, i.e. discrete label prediction.
With the capability of modeling various forms of information, the general-purpose foundation model can be transferred and employed to handle a wide range of practical tasks.

Despite these advantages and great successes in various domains, the application of discrete label prediction in general audio processing remains challenging for two reasons. Firstly, as the audio signal is continuous and the same acoustic event might have various durations in different occasions, it is not straightforward to directly split the audio into semantically meaningful tokens as in language processing \citep{devlin2019bert}. 
On the other hand, different from speech, the general audio signals contain excessively larger data variations, including various non-speech acoustic events and environmental sounds, where the commonly used speech tokenizer for phoneme information extraction \citep{hsu2021hubert} can not be directly applied.

\begin{wrapfigure}{l}{0.4\textwidth}
\includegraphics[width=1.0\linewidth]{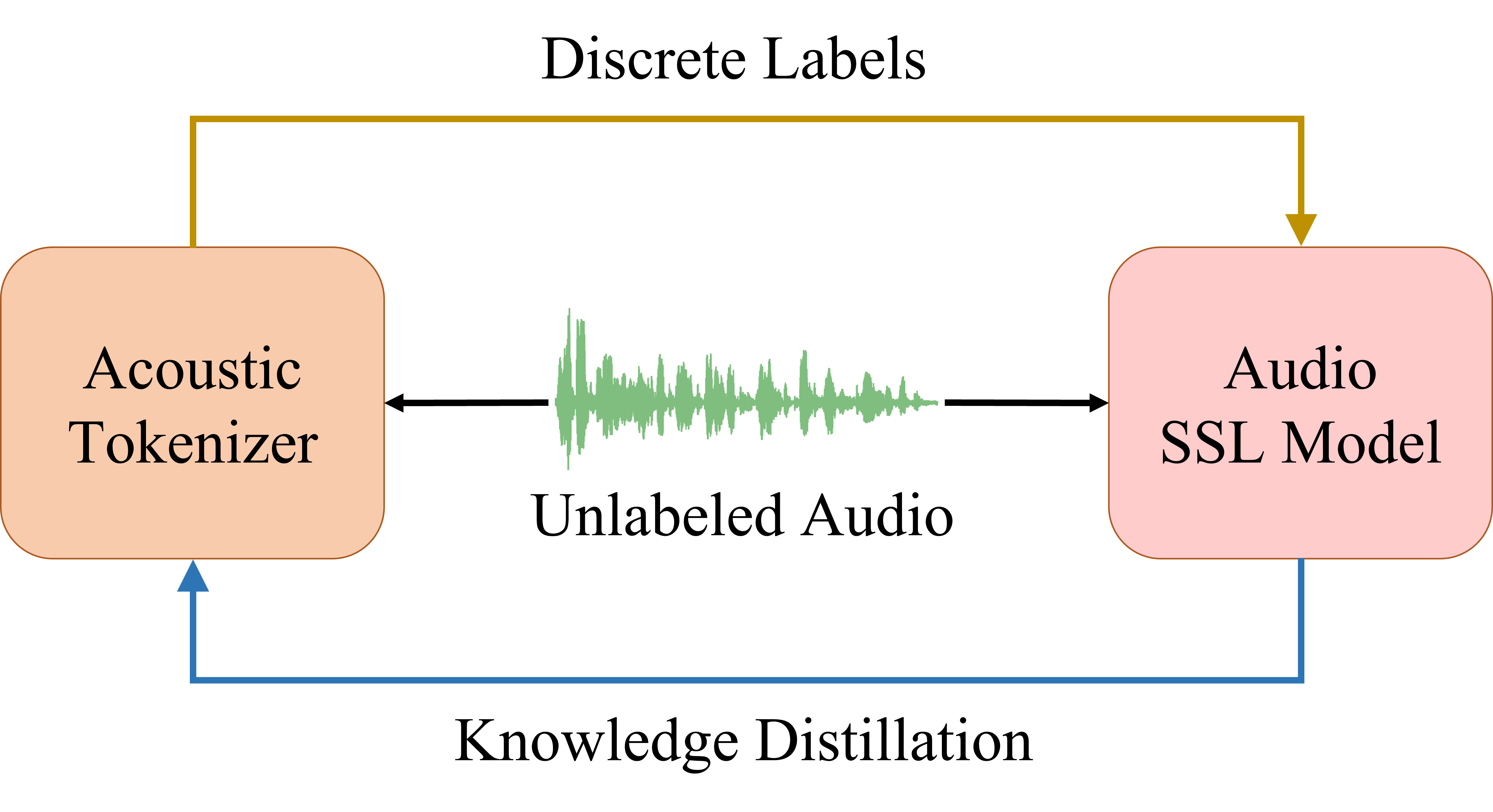} 
\caption{Iterative audio pre-training of \our{}.}
\label{fig:iteration}
\end{wrapfigure}

To tackle these challenges, in this work we propose \our{}, short for  \textbf{B}idirectional \textbf{E}ncoder representation from \textbf{A}udio \textbf{T}ransformer\textbf{s},
in which an acoustic tokenizer and an audio SSL model are optimized through an iterative audio pre-training framework.
The training pipeline is illustrated in Figure \ref{fig:iteration}.
In each iteration, we first use the acoustic tokenizer to generate the discrete labels of the unlabeled audio, and use them to optimize the audio SSL model with a mask and discrete label prediction loss.
After convergence, the audio SSL model acts as a teacher to guide the acoustic tokenizer to learn audio semantics with knowledge distillation \citep{hinton2015distilling}.
In this alternating update learning process, the acoustic tokenizer and the audio SSL model can benefit from each other. The procedure is repeated until convergence.  
Specifically, in the first iteration, we use a random-projection acoustic tokenizer to generate discrete labels as a cold start. 
In addition, we could fine-tune the audio SSL model with a little supervised data, and use the fine-tuned model as the teacher for acoustic tokenizer training. A fine-tuned model learns semantic knowledge not only from SSL but supervised learning, which can further improve the  tokenizer quality. 
We believe the proposed pre-training framework encourages our audio SSL model to learn relevant semantic information from iterations. 
Our pre-training framework is also compatible with any masked audio prediction model, regardless of what backbone network is used. 

We employ the vanilla ViT model \citep{vit} as the backbone of our audio SSL models without heavy structure engineering, and apply the speed-up technique proposed in \citet{he2022masked}. Given the discrete labels generated by the acoustic tokenizer, we mask 75$\%$ of the input sequence and let the model predict the corresponding discrete labels on mask regions. We follow \citet{xu2022masked} to fine-tune the audio SSL model across various audio tasks. 
Experimental results show that our \our{}  pre-trained models have superior performance compared with previous works across six audio and speech classification tasks.
We achieve SOTA audio understanding performance on AudioSet-2M, and outperform the previous SOTA results by a large margin (48.6 v.s. 47.4 for single model and 50.6 v.s. 49.6 for ensemble models) with much fewer model parameters and training data. 
On ESC-50, our \our{} achieved 25\% relative  error rate reduction over the previous SOTA performance.
We further demonstrate the effectiveness of our proposed acoustic tokenizers, where the generated discrete labels are robust to random disturbances and well aligned with audio semantics.

Our contributions include the following:
1) We propose an iterative audio pre-training framework, which opens the door to audio pre-training with a discrete label prediction loss and shows better performance than with reconstruction loss. It unifies the pre-training for speech and audio, which sheds light on the foundation model building for both speech and audio. 
2)  We provide effective acoustic tokenizers to quantize continuous audio features into semantic-rich discrete labels, facilitating future work of audio pre-training and multi-modality pre-training.  
3) We achieve SOTA results on several audio and speech understanding benchmarks. The models and codes are released \footnote{\url{https://aka.ms/beats}}
to facilitate future research.

\section{Related Work}

Recently, audio pre-training has achieved great success in audio understanding tasks.
The existing audio pre-training methods include supervised pre-training and self-supervised pre-training.

\paragraph{Supervised audio pre-training.}
The supervised audio pre-training methods either leverage out-of-domain supervised data (e.g. ImageNet \citep{deng2009imagenet}) or in-domain supervised audio data (e.g. AudioSet \citep{gemmeke2017audio}) for pre-training.
As for the out-of-domain supervised pre-training, PSLA \citep{gong2021psla}  proposes to use an ImageNet supervised pre-trained EfficientNet \citep{tan2019efficientnet} model for the audio model initialization and fine-tunes the model on the audio classification tasks, which leads to significant accuracy improvement.
Instead of CNNs \citep{lecun1995convolutional}, AST \citep{gong2021ast}, PaSST \citep{koutini2021efficient}, MBT \citep{nagrani2021attention} and HTS-AT \citep{chen2022hts} employ Transformer-based architectures \citep{vaswani2017attention} as the backbone, such as ViT \citep{vit} and Swin Transformer \citep{liu2021swin},  and obtain superior audio classification performance.

As for the in-domain supervised pre-training, inspired by the vision pre-training method CLIP \citep{radford2021learning}, CLAP \citep{elizalde2022clap} proposes a contrastive language-audio pre-training task to learn the text-enhanced audio representations with supervised audio and text pairs.
Instead of pre-training from scratch, Wav2clip \citep{wu2022wav2clip} and Audioclip \citep{guzhov2022audioclip} leverage the CLIP pre-trained model and learn an additional audio encoder with the supervised pairs of audio and class labels from AudioSet.
In addition, to push the performance for audio classification tasks with scarce data, some previous works \citep{kong2020panns,verbitskiy2022eranns,gong2021ast,chen2022hts,koutini2021efficient,xu2022masked} report the results on ESC-50 (1.6K training samples) with an additional round of supervised pre-training on the AudioSet dataset (2M training samples).
Despite the promising classification results, these methods strongly rely on a great amount of supervised data, which is complex and expensive in practice.

\paragraph{Self-supervised audio pre-training.}
In comparison, the self-supervised pre-training methods only require large-scale unlabeled data, which can be easily get from the Internet. 
The self-supervised audio pre-training methods typically learn the audio representations with the contrastive learning or reconstruction objective.
LIM \citep{ravanelli2018learning}, COLA \citep{saeed2021contrastive} and \citet{fonseca2021unsupervised}  adopt the contrastive learning framework for audio pre-training, where the positive samples are the augmented clips from the same audio, and the negative ones are sampled from the different audios. 
Instead of taking only the raw waveform or the acoustic feature as the input, CLAR \citep{al2021clar} proposes several data augmentation methods on both of them for more effective contrastive learning. \citet{wang2021multi} also propose a contrastive learning framework with different formats of audio samples by maximizing the agreement between the
raw waveform and its acoustic feature.

As for the reconstruction pre-training objective, inspired by Word2Vec \citep{Word2Vec} in NLP, Audio2Vec \citep{tagliasacchi2020pre} proposes the CBoW task to reconstruct the acoustic feature of an audio clip of pre-determined duration 
based on past and future clips, and the skip-gram task to predict the past and future clips based on the middle audio clip.
BYOL-A \citep{niizumi2021byol} adopts the siamese architecture as BYOL \citep{grill2020bootstrap}, and learns to encode the robust audio representations that are invariant to different audio augmentation methods with the mean square error (MSE) loss and exponential moving average (EMA) optimization strategy.
SSAST \citep{gong2022ssast} proposes a patch-based self-supervised learning method to pre-train AST \citep{gong2021ast} with both the reconstruction and contrastive loss,
and obtains comparable performance to the supervised pre-training methods.
Inspired by the success of the recent visual pre-training method MAE \citep{he2022masked}, 
MSM-MAE \citep{niizumi2022masked}, MaskSpec \citep{chong2022masked}, MAE-AST \citep{baade2022mae} and Audio-MAE \citep{xu2022masked} learn the audio representations following the Transformer-based encoder-decoder design and reconstruction pre-training task in MAE, where a decoder is trained to reconstruct the masked patches based on the encoded representations of the unmasked ones.
Until now, the MAE-style reconstruction pre-training methods show the best audio understanding performance on various audio classification tasks.

In addition, Audio2Vec \citep{tagliasacchi2020pre} proposes the TemporalGap pre-training task to estimate the absolute time distance between two audio clips, which is however inferior to the reconstruction tasks.
\citet{carr2021self} introduces a permutation-based self-supervised pre-training method, where the model is trained to reorder the shuffled patches of an input acoustic feature, and leverage differentiable ranking to enable end-to-end model pre-training.
Unlike the previous methods, in this work, we explore the self-supervised audio pre-training method with the masked discrete label prediction objective for the first time.

\paragraph{Audio and speech tokenizer.} 
Various tokenizers have been proposed for learning discrete representations on audio and speech tasks. 
\cite{dieleman2018challenge} propose a hierarchical VQ-VAE based model to learn audio discrete representations for music generation tasks. 
HuBERT \citep{hsu2021hubert} generates discrete labels with the iterative hidden state clustering method for speech SSL task, where the hidden state is extracted from the last round speech SSL model.
\cite{chiu2022self} claim a random-projection tokenizer is adequate for a large speech SSL model pre-training.
Our work is the first to train an acoustic tokenizer with the supervision of the last round SSL model,  which is different from the previous auto-encoding and ad-hoc clustering methods.  

\section{\our{}}

\subsection{Iterative Audio Pre-training}

Figure \ref{fig:iteration} shows the overall pipeline of our iterative audio pre-training framework of \our{}, where an acoustic tokenizer (Section~\ref{ssec:tokenizer}) and an audio SSL model (Section~\ref{ssec:sslmodel}) are optimized by iterations.
In each iteration, given the unlabeled audio, we use the acoustic tokenizer to generate the discrete labels, and use them to train the audio SSL model with a mask and discrete label prediction loss.
After model convergence, we use the audio SSL model as the teacher to train a new acoustic tokenizer with knowledge distillation for the next iteration of audio SSL model training.

Specifically, given an audio clip as the input, we first extract the corresponding acoustic features, split them into regular grid patches, and further flatten them to the patch sequence $\mathbf{X} = \{\mathbf{x}_t\}_{t=1}^T$. 
For the audio SSL model training, we use the acoustic tokenizer to quantize the patch sequence $\mathbf{X}$ to the patch-level discrete labels $\hat{Z} = \{\hat{z}_t\}_{t=1}^T$ as the masked prediction targets.
For the acoustic tokenizer training, we leverage the audio SSL model to encode the patch sequence $\mathbf{X}$ and extract the output sequence  $\mathbf{\hat{O}} = \{\mathbf{\hat{o}}_{t}\}_{t=1}^T$ as the knowledge distillation targets.
 
Note that we could leverage either a pre-trained audio SSL model or a fine-tuned audio SSL model as the teacher for acoustic tokenizer training.
A fine-tuned model learns semantic knowledge not only from self-supervised pre-training but supervised fine-tuning, making it a better teacher for audio semantics distillation.
With this alternating update learning process, the acoustic tokenizer benefits from the semantic-rich knowledge encoded by the audio SSL model, while the audio SSL model benefits from semantic-rich discrete labels generated by the acoustic tokenizer.
The procedure is repeated until convergence. 

\subsection{Acoustic Tokenizers}
\label{ssec:tokenizer}
The acoustic tokenizers are used to generate the discrete labels for each iteration of \our{} pre-training.
In the first iteration, given the teacher model is unavailable, we employ a Random-Projection Tokenizer (Section~\ref{sssec:random_tokenizer}) to cluster the continuous acoustic features into discrete labels as a cold start.
Starting from the second iteration, we train a Self-Distilled Tokenizer (Section~\ref{sssec:distill_tokenizer}) to generate the refined discrete labels with the semantic-aware knowledge distilled from the pre-trained/fine-tuned audio SSL model obtained in the last iteration. 
\begin{figure*}[t]
	\centering
	\includegraphics[width=\textwidth]{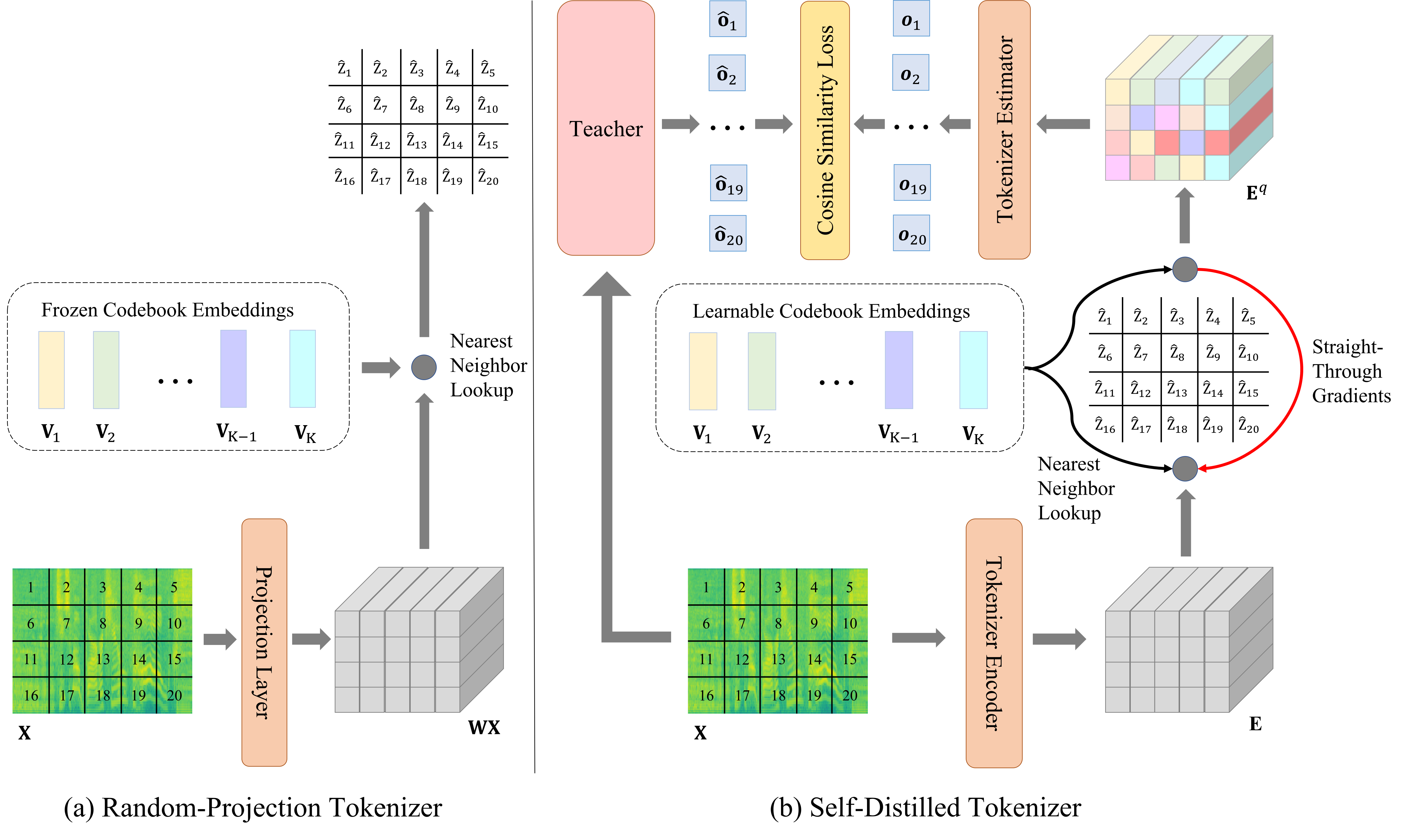}
	\caption{Acoustic tokenizers for discrete label generation.}\label{tokenizer}
\end{figure*}

\subsubsection{Cold Start: Random-Projection Tokenizer}
\label{sssec:random_tokenizer}

For the first iteration of \our{}  pre-training, we apply the random-projection tokenizer \citep{chiu2022self} to generate the patch-level discrete labels for each input audio.

As shown in the left part of Figure~\ref{tokenizer}, the random-projection tokenizer includes a linear projection layer and a set of codebook embeddings, which are kept frozen after random initialization.
Each patch of the input feature is first projected with the linear layer, then finds the nearest neighbor vector among the codebook embeddings, where the index of the nearest neighbor is defined as the discrete label.

Specifically, given the patch sequence extracted from the input audio $\mathbf{X} = \{\mathbf{x}_t\}_{t=1}^T$, we first project $\mathbf{x}_t$ to the vector $\mathbf{W} \mathbf{x}_t$ with a randomly initialized projection layer $\mathbf{W}$.
Then we look up the nearest neighbor vector of each projected vector $\mathbf{W} \mathbf{x}_t$ from a set of random initialized vectors $\mathbf{V} = \{\mathbf{v}_i\}_{i=1}^K$, where $K$ is the codebook size, and define the discrete label of $t$-th patch as the index of the nearest neighbor vector: 
\begin{equation}
    \hat{z}_t = \argmin_i || \mathbf{v}_i - \mathbf{W} \mathbf{x}_t ||_2^2
\end{equation}

\subsubsection{Iteration: Self-Distilled Tokenizer}
\label{sssec:distill_tokenizer}

From the second iteration of \our{}  pre-training, we leverage the last iteration audio SSL  model as the teacher, which can be either a pre-trained model or a fine-tuned model, to teach the current iteration tokenizer learning. We call it the self-distilled tokenizer to generate the patch-level discrete labels for each input audio.

As shown in the right part of Figure~\ref{tokenizer}, the self-distilled tokenizer first uses a Transformer-based tokenizer encoder to convert the input patches to discrete labels with a set of learnable codebook embeddings.
Then, a Transformer-based tokenizer estimator is trained to predict the output of a  teacher model with the discrete labels and codebook embeddings as the input. 
With knowledge distillation as the training target, the tokenized discrete labels are optimized to contain more  semantic-rich knowledge from the teacher and less  redundant information of the input audio.

Specifically, we first feed the input patches $\mathbf{X} = \{\mathbf{x}_t\}_{t=1}^T$  to a 12-layer Transformer encoder and obtain the encoded vector sequence $\mathbf{E} = \{\mathbf{e}_t\}_{t=1}^T$. 
Then, for each encoded vector $\mathbf{e}_t$, we conduct the quantization by finding the nearest neighbor vector $\mathbf{v}_{\hat{z}_t}$ from the codebook embeddings $\mathbf{V} = \{\mathbf{v}_i\}_{i=1}^K$:
\begin{equation}
\label{eq:quantize}
    \hat{z}_t = \argmin_i || \ell_2( \mathbf{v}_i) - \ell_2(\mathbf{e}_t) ||_2^2,
\end{equation}
where $\ell_2$ normalization is used to improve the codebook utilization \citep{yu2021vector,beitv2}. 
With the quantized vector sequence $\mathbf{E}^q = \{\mathbf{v}_{\hat{z}_t}\}_{t=1}^T$ as the input, we use a 3-layer Transformer estimator to predict the last layer output of the teacher model $\{\mathbf{\hat{o}}_{t}\}_{t=1}^T$.

To deal with the non-differentiable problem of the vector quantization (Equation~\ref{eq:quantize}), following \citet{van2017neural}, we apply the straight-through gradients mechanism, where the gradients are directly copied from the quantized vector sequence $\mathbf{E}^q$ to the encoded vector sequence $\mathbf{E}$ during the backward process. 

The overall training objective of the self-distilled tokenizer is defined as the cosine similarity between the output sequence of the tokenizer estimator $\{\mathbf{o}_{t}\}_{t=1}^T$ and the output sequence of the teacher model $\{\mathbf{\hat{o}}_{t}\}_{t=1}^T$, along with the mean squared error between the encoded vector sequence $\mathbf{E}=\{\mathbf{e}_t\}_{t=1}^T$ and the quantized vector sequence $\mathbf{E}^q=\{\mathbf{v}_{\hat{z}_t}\}_{t=1}^T$:
\begin{equation}
    \max \sum_{\mathbf{X} \in \mathcal{D}} \sum_{t=1}^T \cos (\mathbf{o}_{t}, \mathbf{\hat{o}}_{t}) - || sg[\ell_2(\mathbf{e}_t)] - \ell_2( \mathbf{v}_{\hat{z}_t}) ||_2^2 - || \ell_2(\mathbf{e}_t) - sg[\ell_2( \mathbf{v}_{\hat{z}_t})] ||_2^2,
\end{equation}
where $\mathcal{D}$ denotes the pre-training datasets, $\cos(\cdot, \cdot)$ and $sg[\cdot]$ are the cosine similarity and the stopgradient operator, respectively. We  employ the exponential moving average \citep{van2017neural} for codebook embedding optimization for more stable tokenizer training \citep{beitv2}.

During inference, we discard the tokenizer estimator, and leverage the pre-trained tokenizer encoder and codebook embeddings to convert each input audio $\mathbf{X} = \{\mathbf{x}_t\}_{t=1}^T$ to patch-level discrete labels $\hat{Z} = \{\hat{z}_t\}_{t=1}^T$, as in Equation~\ref{eq:quantize}.

\subsection{Audio SSL Model}
\label{ssec:sslmodel}

\begin{figure*}[t]
	\centering
	\includegraphics[width=\textwidth]{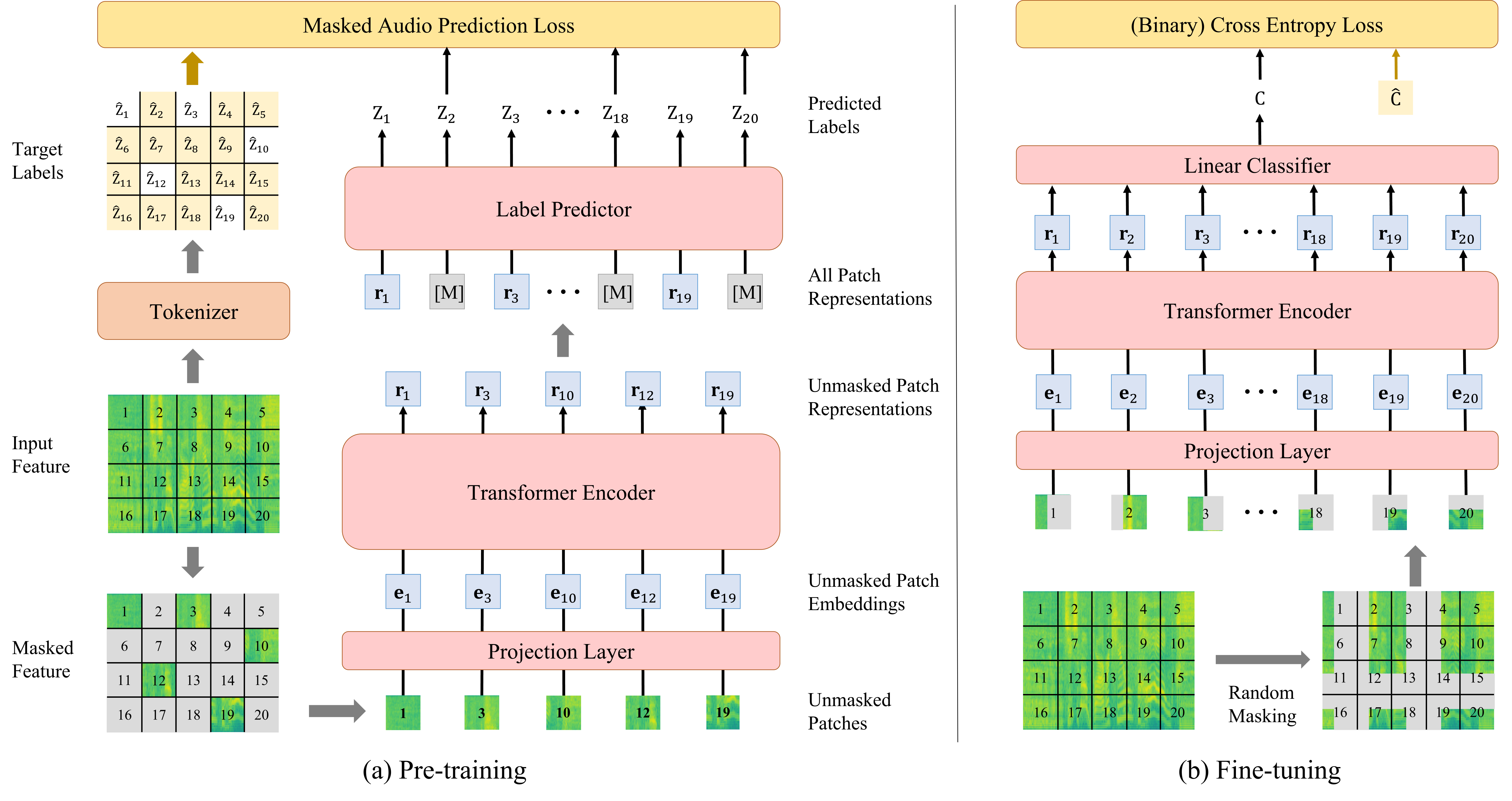}
	\caption{Overview of audio SSL model pre-training and fine-tuning.}\label{fig:architecture}
\end{figure*}

\subsubsection{Backbone}

Following the previous works \citep{gong2021ast,gong2022ssast,xu2022masked}, we employ the ViT structure \citep{vit} as the backbone network, which consists of a linear projection layer and a stack of Transformer encoder layers.

Given the patch sequence extracted from the input audio $\mathbf{X} = \{\mathbf{x}_t\}_{t=1}^T$, we first convert them to the patch embeddings $\mathbf{E} = \{\mathbf{e}_t\}_{t=1}^T$ with a linear projection network.
Then, we feed the patch embeddings to the Transformer encoder layers, and obtain the encoded patch representations $\mathbf{R} = \{\mathbf{r}_t\}_{t=1}^T$. The Transformer is equipped with a convolution-based relative position embedding layer at the bottom, and the gated relative position
bias \citep{chi2022xlm} for better position information encoding. We also employ the DeepNorm \citep{wang2022deepnet} for more stable pre-training.

\subsubsection{Pre-Training}

We propose a Masked Audio Modeling (MAM) task for the audio SSL model pre-training, as shown in the left part of Figure~\ref{fig:architecture}.
Different from the previous audio pre-training methods, where the model is optimized to reconstruct the input acoustic feature, our model 
 is optimized to predict the patch-level discrete labels generated by the acoustic tokenizers (Section~\ref{ssec:tokenizer}) with a Transformer-based label predictor. 

Specifically, given the input patch sequence $\mathbf{X} = \{\mathbf{x}_t\}_{t=1}^T$ and the corresponding target discrete acoustic labels $\hat{Z} = \{\hat{z}_t\}_{t=1}^T$, we randomly mask 75\% of the input patches, where the masked positions are denoted as $\mathcal{M} = \{1, \dots, T\}^{0.75T}$.
Then, we feed the unmasked patch sequence $\mathbf{X}^{U} = \{\mathbf{x}_t : t \in \mathcal{M}\}_{t=1}^T$ to the ViT encoder, and obtain the encoded representations $\mathbf{R}^{U} = \{\mathbf{r}_t : t \in \mathcal{M}\}_{t=1}^T$.
Finally, we feed the combination of the non-masked patch representations and the masked patch features $\{\mathbf{r}_t : t \in \mathcal{M}\}_{t=1}^T \cup \{\mathbf{0} : t \not\in \mathcal{M}\}_{t=1}^T$ to the label predictor to predict the discrete acoustic labels $Z = \{z_t\}_{t=1}^T$. It should be noted here that only feeding the non-masked patches into the encoder could significantly speed up the training process while providing slight improvement across downstream tasks \citep{xu2022masked}.

The pre-training objective of MAM is the cross entropy loss which maximizes the log-likelihood of the correct acoustic labels in the masked positions given the unmasked patch sequences.
\begin{equation}
    \mathcal{L}_{\text{MAM}} = -\sum_{t \in \mathcal{M}} \log p(\hat{z}_t|\mathbf{X}^{U})
\end{equation}

\subsubsection{Fine-Tuning}

During audio SSL model fine-tuning, we discard the label predictor, and append a task-specific linear classifier upon the ViT encoder to generate the labels for the downstream classification tasks, as shown in the right part of Figure~\ref{fig:architecture}.

Specifically, we first random mask the input acoustic feature in the time and frequency dimension as spec-augmentation \citep{park2019specaugment}, then split and flat it to the patch sequence $\mathbf{X} = \{\mathbf{x}_t\}_{t=1}^T$.
Unlike pre-training, we feed the whole patch sequence $\mathbf{X}$ to the ViT encoder, and obtain the encoded representations $\mathbf{R} = \{\mathbf{r}_t\}_{t=1}^T$.
Finally, we use a linear classifier to calculate the category probabilities as $p(C) = \text{Softmax} ( \text{MeanPool}(\mathbf{W}_c\mathbf{R} ))$, where Softmax, MeanPool and $\mathbf{W}_c$ denote the softmax operation, mean-pooling layer and the linear projection, respectively. 

We employ the cross entropy loss as the fine-tuning objective for the single label classification tasks, and the binary cross entropy loss for the multi-label classification tasks or the mixup augmentation \citep{zhang2017mixup} is employed.

\section{Experiment}

\subsection{Datasets}

We pre-train our \our{}  tokenizers and audio SSL models on the full training set of the AudioSet dataset, and evaluate our pre-trained audio SSL models on six downstream tasks, including three audio classification tasks (AS-2M, AS-20K and ESC-50) and three speech classification tasks (KS1, KS2 and ER). 

\textbf{AudioSet (AS-2M and AS-20K)} \citep{gemmeke2017audio} is a large-scale audio classification dataset. It contains over 2 million 10-second YouTube clips annotated with 527 audio event classes, where each clip could be annotated with multiple audio event classes. It is officially subdivided into three partitions, including a class-wise balanced set (22,176 clips), a class-wise unbalanced set (2,042,985 clips), and an eval set (20,383 clips). Due to the constant change in YouTube video availability (e.g., videos being removed or taken down), we downloaded and processed 20,666, 1,919,153, and 18,987 clips for the balanced, unbalanced, and eval sets, respectively, which is consistent with the previous works \citep{baade2022mae}.

Following the previous works, we use the combination of the 21K balanced and the 1.9M unbalanced training audios for fine-tuning in the AS-2M task, and only the 21K balanced training audios for  fine-tuning in the AS-20K task. We evaluate our models on the 19K eval set with the  mean average precision (mAP) evaluation metric.

\textbf{Environmental Sound Classification (ESC-50)} \citep{piczak2015esc} is an audio classification dataset that contains 2,000 5-second environmental sound recordings annotated with 50 classes. Each sound recording is only annotated with one class. We follow the 5-fold cross-validation evaluation setting as the previous works and report the classification accuracy as the evaluation metric.

\textbf{Speech Commands V2 (KS2)} \citep{warden2018speech} is a keyword spotting dataset that contains  105,829 1-second spoken word clips annotated with 35 common word classes. It is officially subdivided into the training, validation, and testing set that contains 84,843, 9,981, and 11,005 audio clips respectively. We report classification accuracy as the evaluation metric.

\textbf{Speech Commands V1 (KS1)} \citep{warden2018speech} task uses the same dataset as KS2, but only contains 10 classes of keywords, 1 silence class, and 1 unknown class that includes all the other 20 common speech commands. We use the standard data and split provided in SUPERB benchmark \citep{SUPERB} to report classification accuracy for a fair comparison with the previous works.

\textbf{IEMOCAP (ER)} \citep{busso2008iemocap} is an emotion recognition dataset that contains about 12 hours of emotional speech clips annotated with four classes. we use the 5-fold cross-validation evaluation setting as SUPERB benchmark \citep{SUPERB} and report classification accuracy as the evaluation metric.

\subsection{Implementation Details}

\paragraph{Backbone}
The \our{} models have 12 Transformer encoder layers, 768-dimensional hidden states, and 8 attention heads, resulting in 90M parameters.
We keep the model size similar to the previous SOTA audio pre-trained models \citep{xu2022masked,chong2022masked} for a fair comparison of the pre-training methods.

\paragraph{Acoustic feature}
Following \citep{gong2021ast,gong2022ssast}, we convert the sample rate of each raw waveform to 16,000, and extract the 128-dimensional Mel-filter bank features with a 25ms Povey window that shifts every 10 ms as the acoustic feature. 
The acoustic feature is normalized to the mean value of 0 and standard deviation of 0.5 following the previous works.
We split each acoustic feature into the 16  $\times$ 16 patches, and further flat them to the patch sequence as the input of our \our{} tokenizers and models.

\paragraph{Model and tokenizer training}
We pre-train the \our{} models on AS-2M dataset for three iterations and denote them as \our{}$_\text{iter1}$, \our{}$_\text{iter2}$, \our{}$_\text{iter3}$, \our{}$_\text{iter3+}$.

The \our{}$_\text{iter1}$ is pre-trained with the discrete labels generated by a random-projection tokenizer (Section~\ref{sssec:random_tokenizer}). 
Starting from the second iteration, we train a self-distilled tokenizer (Section~\ref{sssec:distill_tokenizer}) to generate the discrete labels for the pre-training of \our{}$_\text{iter2}$ and \our{}$_\text{iter3}$ with the pre-trained \our{}$_\text{iter1}$ and \our{}$_\text{iter2}$ as the teacher, respectively.
Different from \our{}$_\text{iter3}$, the self-distilled tokenizer for \our{}$_\text{iter3+}$ pre-training takes the supervised fine-tuned \our{}$_\text{iter2}$ as the teacher model and learns to estimate the classification logits of the input audios.
Compared with the other \our{} models, the \our{}$_\text{iter3+}$ not only make use of the downstream supervised data during fine-tuning but also in pre-training.

We pre-train all the \our{} models for 400k steps with a batch size of 5.6K seconds and a 5e-4 peak learning rate.
The codebook of all the tokenizers contains 1024 embeddings with 256 dimensions.
The self-distilled tokenizer with a self-supervised model as the teacher is trained for 400k steps with a batch size of 1.4K seconds and a 5e-5 peak learning rate.
The self-distilled tokenizer with a supervised model as the teacher is trained for 400k steps with a batch size of 1.4K seconds and a 5e-4 peak learning rate.
Please see Appendix~\ref{ssec:hyper} for the detailed hyperparameter settings.

\subsection{Comparing with the SOTA Single Models}

\begin{table*}[t]
    \centering
    \caption{Comparing with the SOTA single models on audio and speech classification tasks. IN, AS, and LS denote the ImageNet, AudioSet, and LibriSpeech datasets, respectively. TA and TI denote the 128K text-audio pairs and 400M text-image pairs for CLAP and CLIP pre-training, respectively. The evaluation metrics are mAP for AS-2M/AS-20K and accuracy for ESC-50/KS1/KS2/ER. We compared the best single models from each previous work. 
    We \textcolor{gray}{gray-out} the models and results with additional supervised training on the external datasets.
    $^*$The results reported following the SUPERB policy \citep{SUPERB}, where pre-trained models are kept frozen during fine-tuning.
    }
    \label{tab:main}
    \resizebox{\columnwidth}{!}{
    \begin{tabular}{lllllllll} 
         \toprule 
         \multirow{2}{*}{Model}
         & \multirow{2}{*}{\# Param} 
         & \multirow{2}{*}{Data} 
         & \multicolumn{3}{c}{Audio}
         & \multicolumn{3}{c}{Speech} \\
				\cmidrule(lr){4-6}
				\cmidrule(lr){7-9}
         & & & AS-2M & AS-20K & ESC-50 & KS1 & KS2 & ER \\
         \hline
         \multicolumn{8}{l}{\textbf{{No Pre-Training}}} \\
         PANN \citep{kong2020panns} & 81M & - & 43.1 & 27.8 & 83.3  & - & 61.8 & -  \\
         ERANN \citep{verbitskiy2022eranns} & 55M & - & 45.0 & - & 89.2  & - & - & - \\
         \hline
         \multicolumn{8}{l}{\textbf{{Out-of-domain Supervised Pre-Training}}} \\
         \rowstyle{\color{gray}}
         \textcolor{gray}{PSLA} \citep{gong2021psla} & \textcolor{gray}{14M} & \textcolor{gray}{IN} &  \textcolor{gray}{44.4} & \textcolor{gray}{31.9} & \textcolor{gray}{}- & \textcolor{gray}{-} & \textcolor{gray}{96.3} & \textcolor{gray}{-} \\
         \textcolor{gray}{AST} \citep{gong2021ast} & \textcolor{gray}{86M} & \textcolor{gray}{IN} &  \textcolor{gray}{45.9} & \textcolor{gray}{34.7} & \textcolor{gray}{88.7} & \textcolor{gray}{95.5} & \textcolor{gray}{98.1} & \textcolor{gray}{56.0} \\
         \textcolor{gray}{MBT} \citep{nagrani2021attention} & \textcolor{gray}{86M} & \textcolor{gray}{IN-21K} &  \textcolor{gray}{44.3} & \textcolor{gray}{31.3} & \textcolor{gray}{}- & \textcolor{gray}{-} & \textcolor{gray}{-} & \textcolor{gray}{-} \\
         \textcolor{gray}{PaSST} \citep{koutini2021efficient} & \textcolor{gray}{86M} & \textcolor{gray}{IN} &  \textcolor{gray}{47.1}  & \textcolor{gray}{-} & \textcolor{gray}{-} & \textcolor{gray}{-} & \textcolor{gray}{-} & \textcolor{gray}{-} \\
         \textcolor{gray}{HTS-AT} \citep{chen2022hts} & \textcolor{gray}{31M} & \textcolor{gray}{IN} &  \textcolor{gray}{47.1} & \textcolor{gray}{-} & \textcolor{gray}{-} & \textcolor{gray}{-} & \textcolor{gray}{98.0} & \textcolor{gray}{-} \\
         \textcolor{gray}{Wav2CLIP} \citep{wu2022wav2clip} & \textcolor{gray}{74M} & \textcolor{gray}{TI+AS} &  \textcolor{gray}{-} & \textcolor{gray}{-} & \textcolor{gray}{86.0} & \textcolor{gray}{-} & \textcolor{gray}{-} & \textcolor{gray}{-} \\
         \textcolor{gray}{AudioCLIP} \citep{guzhov2022audioclip} & \textcolor{gray}{93M} & \textcolor{gray}{TI+AS} &  \textcolor{gray}{25.9} & \textcolor{gray}{-} & \textcolor{gray}{96.7} & \textcolor{gray}{-} & \textcolor{gray}{-} & \textcolor{gray}{-} \\

         \hline
         \multicolumn{8}{l}{\textbf{{In-domain Supervised Pre-Training}}} \\
         \rowstyle{\color{gray}}
         \textcolor{gray}{PANN} \citep{kong2020panns} & \textcolor{gray}{81M} & \textcolor{gray}{AS} &  \textcolor{gray}{-} & \textcolor{gray}{-} & \textcolor{gray}{94.7} & \textcolor{gray}{-} & \textcolor{gray}{-} & \textcolor{gray}{-} \\
         \textcolor{gray}{ERANN} \citep{verbitskiy2022eranns} & \textcolor{gray}{55M} & \textcolor{gray}{AS} &  \textcolor{gray}{-} & \textcolor{gray}{-} & \textcolor{gray}{96.1} & \textcolor{gray}{-} & \textcolor{gray}{-} & \textcolor{gray}{-} \\
         \textcolor{gray}{AST} \citep{gong2021ast} & \textcolor{gray}{86M} & \textcolor{gray}{IN+AS} &  \textcolor{gray}{45.9} & \textcolor{gray}{-} & \textcolor{gray}{95.6} & \textcolor{gray}{-} & \textcolor{gray}{97.9} & \textcolor{gray}{-} \\
         \textcolor{gray}{PaSST} \citep{koutini2021efficient} & \textcolor{gray}{86M} & \textcolor{gray}{IN+AS} &  \textcolor{gray}{47.1} & \textcolor{gray}{-} & \textcolor{gray}{96.8} & \textcolor{gray}{-} & \textcolor{gray}{-} & \textcolor{gray}{-} \\
         \textcolor{gray}{HTS-AT} \citep{chen2022hts} & \textcolor{gray}{31M} & \textcolor{gray}{IN+AS} &  \textcolor{gray}{47.1} & \textcolor{gray}{-} & \textcolor{gray}{97.0} & \textcolor{gray}{-} & \textcolor{gray}{-} & \textcolor{gray}{-} \\
         \textcolor{gray}{CLAP} \citep{elizalde2022clap} & \textcolor{gray}{190.8M} & \textcolor{gray}{TA} &  \textcolor{gray}{-} & \textcolor{gray}{-} & \textcolor{gray}{96.7} & \textcolor{gray}{-} & \textcolor{gray}{96.8} & \textcolor{gray}{-} \\
         \textcolor{gray}{Audio-MAE} \citep{xu2022masked} & \textcolor{gray}{86M} & \textcolor{gray}{AS} &  \textcolor{gray}{-} & \textcolor{gray}{-} & \textcolor{gray}{97.4} & \textcolor{gray}{-} & \textcolor{gray}{-} & \textcolor{gray}{-} \\

         \hline
         \multicolumn{8}{l}{\textbf{{Self-Supervised Pre-Training}}} \\
         Wav2vec \citep{schneider2019wav2vec} & 33M & LS & - & - & - &  96.2 & - &  59.8 \\
         Wav2vec 2.0 \citep{baevski2020wav2vec} & 95M & LS & - & - & - &  96.2$^*$ & - &  63.4$^*$ \\
         SS-AST \citep{gong2022ssast} & 89M & AS+LS & - & 31.0 & 88.8 &  96.0 & 98.0 &  59.6 \\
         MSM-MAE \citep{niizumi2022masked} & 86M & AS & - & - &  85.6 &  - & 87.3 & - \\
         MaskSpec \citep{chong2022masked} & 86M & AS & 47.1 & 32.3 & 89.6 &  - & 97.7 & - \\
         MAE-AST \citep{baade2022mae} & 86M & AS+LS & - & 30.6 & 90.0 &  95.8 & 97.9 & 59.8 \\
         Audio-MAE \citep{xu2022masked} & 86M & AS & 47.3 & 37.1 & 94.1  &  96.9 & \textbf{98.3} & - \\
         data2vec \citep{baevski2022data2vec} & 94M & AS & - &  34.5 & - &  - & - & - \\
         Audio-MAE Large \citep{xu2022masked} & 304M & AS & 47.4 & 37.6 & - &  - & - & - \\
         CAV-MAE \citep{gong2022contrastive} & 86M & AS+IN & 44.9 & 34.2 & - &  - & - & - \\
         \hline
         \multicolumn{7}{l}{\textbf{{Ours}}} \\
         \our{}$_\text{iter1}$ & 90M & AS & 47.9 & 36.0 & 94.0 &  \textbf{98.0} & \textbf{98.3} & 65.9 \\
         \our{}$_\text{iter2}$ & 90M & AS & 48.1 & 38.3 & 95.1 &  97.7 & \textbf{98.3} & \textbf{66.1} \\
         \our{}$_\text{iter3}$ & 90M & AS & 48.0 & 38.3 & \textbf{95.6} &  97.7 & \textbf{98.3} & 64.5 \\
        \our{}$_\text{iter3+}$ & 90M & AS & \textbf{48.6} & \textbf{38.9} & \textcolor{gray}{98.1} &  \textcolor{gray}{98.1} & \textcolor{gray}{98.1} & \textcolor{gray}{65.0} \\
        
         \bottomrule
       
    \end{tabular}
    }
\end{table*}

Table~\ref{tab:main} shows the comparison of the single-model performance of our \our{} pre-trained models and the previous SOTA models.
For a fair comparison with the previous self-supervised pre-training methods, we report the  \our{}$_\text{iter3+}$ fine-tuning results on AS-2M and AS-20K with the models that are pre-trained with the same supervised dataset as fine-tuning.
On the other tasks, we report the \our{}$_\text{iter3+}$ fine-tuning results with the model that is pre-trained with the AS-2M supervised dataset, and compare them with the previous supervised pre-training methods.
Following \citep{xu2022masked,gong2021ast}, we report the \our{}$_\text{iter3+}$ fine-tuning result on ESC-50 with additional supervised training on AS-2M.

Overall, \our{} achieve the best performance across all six audio and speech classification tasks.
\our{}$_\text{iter3+}$ set a new SOTA  single-model audio understanding performance on AS-2M and AS-20K, and outperform the previous
SOTA results by a large margin (48.6 v.s. 47.4 on AS-2M, and 38.9 v.s. 37.6 on AS-20K) with much fewer model parameters (90M v.s. 304M). Notably, \our{}  also significantly outperform all the previous models that use more out-of-domain or in-domain data for supervised or self-supervised pre-training.
On ESC-50, \our{} successfully reduce the SOTA classification error rate from 5.9\% to 4.4\% without any external supervised data, and from 2.6\% to 1.9\% with external AS-2M supervised data.

As shown in the table, our first iteration model \our{}$_\text{iter1}$ which uses a random-projection tokenizer for label generation can already obtain better performance than previous works on five out of six tasks (AS-2M, ESC-50, KS1, KS2, and ER), which demonstrates the superiority of the discrete label prediction loss  comparing to the reconstruction loss. Pre-trained with the refined labels generated by a self-distilled tokenizer, \our{}$_\text{iter2}$ can achieve further performance improvements, especially on the audio classification tasks. 
With SSL on AS-2M, \our{}$_\text{iter1}$ learns to encode the high-level audio representations with semantic-aware knowledge. Taking \our{}$_\text{iter1}$ as the teacher model, the self-distilled tokenizer is optimized to refine the labels with more audio-related semantics, resulting in the more powerful audio modeling ability of \our{}$_\text{iter2}$.

As for the third iteration of \our{} pre-training, we can find that \our{}$_\text{iter3}$ obtains similar performance as \our{}$_\text{iter2}$, indicating our self-distilled tokenizer is robust to difference SSL teacher models, and our \our{} iterative pre-training procedure is capable of fast convergence in only a few iterations.
Furthermore, if we use the fine-tuned \our{}$_\text{iter2}$ models as the teacher model, the \our{}$_\text{iter3+}$ can bring significant performance gains on both AS-2M and AS-20K tasks, and outperform all the previous SOTA models by a large margin. 
By leveraging the supervised fine-tuning data in our iterative training pipeline, both the acoustic tokenizer and the audio SSL model learn more task-specific semantic knowledge from each other, which would effectively promote  \our{}$_\text{iter3+}$ performance on the downstream understanding tasks.

\subsection{Comparing Different \our{} Tokenizers}
Table~\ref{tab:tokenizer} shows the detailed performance comparison of different \our{} tokenizers.
We can find that the self-distilled tokenizer shows remarkable superiority compared with the random-projection tokenizer, especially in the task with scarce data. 
It is because the random-projection tokenizer with a simple feature clustering process is insufficient to provide the labels with the high-level audio semantic abstraction, while the self-distilled tokenizer is able to distill the semantic knowledge from a well pre-trained audio SSL model to the generated discrete labels. 

In addition, the results show that the performance of the self-distilled tokenizer is insensitive to different self-supervised teachers (e.g. \our{} models) but sensitive to different supervised teachers (e.g. the fine-tuned \our{} models).
The self-distilled tokenizer guided by \our{}$_\text{iter1}$ obtains similar performance as the tokenizer guided by \our{}$_\text{iter2}$.
The self-distilled tokenizer guided by the AS-2M fine-tuned \our{}$_\text{iter2}$ model can achieve the best performance on all three audio classification tasks.

\begin{table*}[t]
    \centering
    \caption{Comparing different \our{} tokenizers on audio classification tasks.
    SSL Data and SL Data denote the training data used for self-supervised learning and supervised learning, respectively.
    $^*$We use AS-2M supervised data during pre-training and AS-20K supervised data during fine-tuning.
    $^\dagger$Here, We report the ESC-50 results without additional supervised pre-training on AS-2M for a fair comparison of different tokenizers.
    }
    \label{tab:tokenizer}
    \resizebox{\columnwidth}{!}{
    \begin{tabular}{lllllccc} 
         \toprule 
         {Model}
         & {Tokenizer Type} 
         & {Tokenizer Teacher} 
         & {SSL Data} 
         & {SL Data} 
         & AS-2M & AS-20K & ESC-50 \\
         \hline
         \our{}$_\text{iter1}$ & Random-Projection & N/A & AS & - & 47.9 & 36.0 & 94   \\
         \our{}$_\text{iter2}$ & Self-Distilled & \our{}$_\text{iter1}$ & AS & - & 48.1 & 38.3 & 95.1  \\
         \our{}$_\text{iter3}$ & Self-Distilled & \our{}$_\text{iter2}$ & AS & - & 48.0 & 38.3 & 95.6  \\
         \our{}$_\text{iter3+}$ & Self-Distilled & \our{}$_\text{iter2}$ fine-tuned on AS-20K & AS & AS-20K & 48.0 & 38.9 & 96.2  \\
         \our{}$_\text{iter3+}$ & Self-Distilled &  \our{}$_\text{iter2}$ fine-tuned on AS-2M & AS & AS & \textbf{48.6} & \textbf{41.8}$^*$ & \textbf{97.1}$^\dagger$  \\
         
         \bottomrule
       
    \end{tabular}
    }
\end{table*}

\begin{figure}[tb]
	\centering
     \subfigure[Reconstruction]{\includegraphics[width=0.28\textwidth]{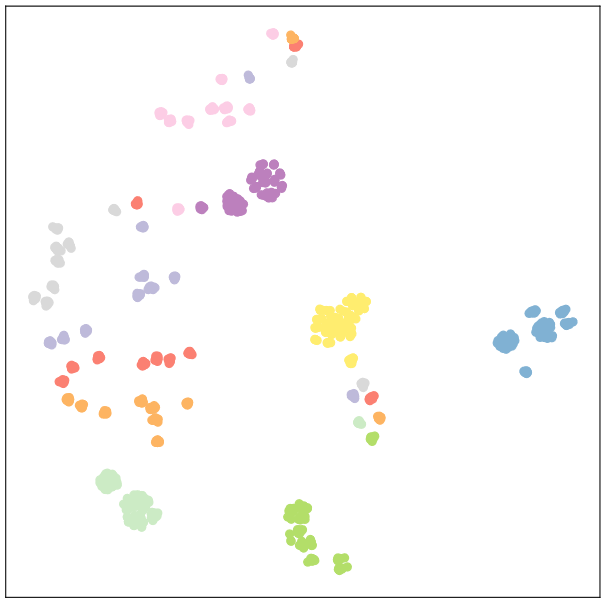}
     \label{fig:visual1}} 
	\subfigure[\our{}$_\text{iter3}$]{\includegraphics[width=0.28\textwidth]
     {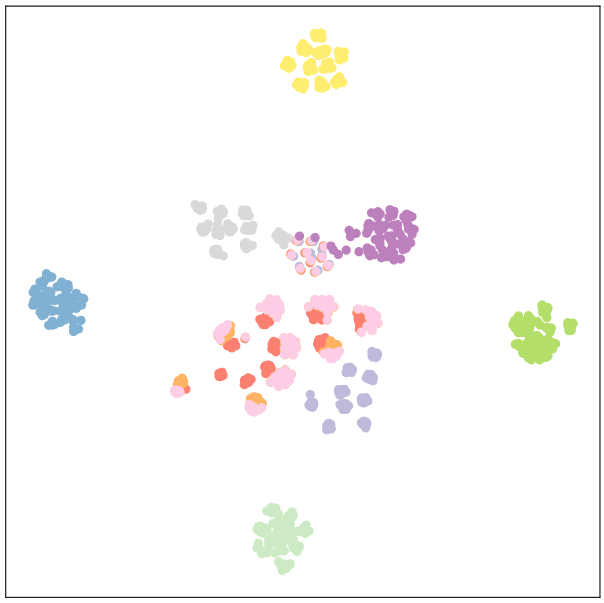}\label{fig:visual2}} 
	\subfigure[\our{}$_\text{iter3+}$]{\includegraphics[width=0.28\textwidth]
    {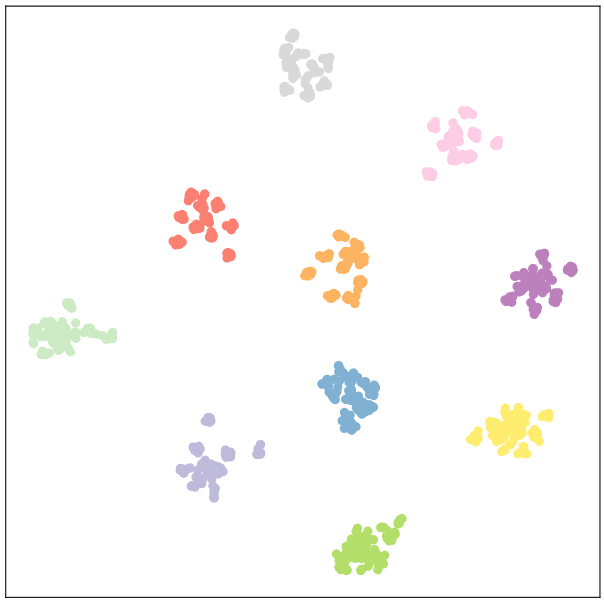}\label{fig:visual3}} 
	\includegraphics[width=0.1\textwidth]{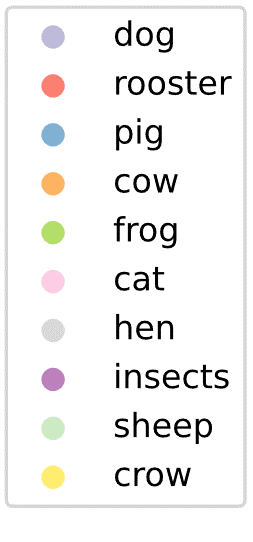} 
	\vspace*{-3mm}
	\caption{Comparing the pre-training targets of different SSL models with audio samples from ESC-50. We visualize the acoustic features for reconstruction-based SSL models, the representations quantized by the tokenizer with a self-supervised pre-trained teacher for \our{}$_\text{iter3}$, and the representations quantized by the tokenizer with a supervised fine-tuned teacher for \our{}$_\text{iter3+}$.}
    \label{fig:visual}
    \centering
\end{figure}

\subsection{Comparing Different Pre-Training Targets via Visualization}

Figure~\ref{fig:visual} shows the comparison of the pre-training targets of different SSL models with audio samples from the ESC-50 dataset.
Specifically, figure~\ref{fig:visual1} demonstrates the acoustic features which are the pre-training targets for reconstruction-based SSL models. Figure~\ref{fig:visual2} and \ref{fig:visual3} illustrate the pre-training targets of \our{}$_\text{iter3}$ and \our{}$_\text{iter3+}$, which are demonstrated with the quantized representations encoded by the acoustic tokenizers with a self-supervised teacher (i.e. \our{}$_\text{iter2}$) and a supervised teacher (i.e. the fine-tuned \our{}$_\text{iter2}$), respectively.
We reduced the feature dimension to 2-D by T-SNE \citep{van2008visualizing} for better visualization.

As the standard evaluation setting, we divide the data into a 1.6K training set and a 0.4K valid set.
We use the training set for \our{}$_\text{iter3+}$ pre-training, and the valid set for visualization.
We randomly select ten audio samples with different classification labels from the valid set, then add some random disturbance on the waveform with RIR \footnote{\url{https://www.openslr.org/28/}} reverberations and DNS noises \citep{reddy2021interspeech}.
The points with different colors denote the audios with different classification labels, and the points with the same color denote different disturbances to the same audio.

As shown in the figures, the pre-training targets of reconstruction-based SSL models are very sensitive to random disturbances on the waveform. 
The acoustic feature of the same audio with different disturbances can be far apart, and the acoustic feature with different labels can be closely spaced.
It indicates the pre-training targets of reconstruction-based SSL models mainly contain low-level time-frequency features and lack high-level audio semantic abstractions.
In comparison, the pre-training targets of \our{} models are much more robust to the random variations. With the self-supervised pre-trained model as the teacher, the acoustic tokenizer learns to cluster the audio samples with the same semantic content and get rid of the background reverberations and noises. With the supervised fine-tuned model as the teacher, the acoustic tokenizer can successfully 
capture high-level semantics of audio regardless of the low-level details of redundancy, and generate semantic-rich discrete tokens for more effective \our{} model pre-training.

\begin{table}[t]
    \centering
    \caption{Comparing with the SOTA ensemble models on AS-2M.
    }
    \label{tab:ensemble}
    \resizebox{0.5\columnwidth}{!}{
    \begin{tabular}{lllccc} 
         \toprule 
         {Model}
         & {SL Data} 
         & AS-2M \\
         \hline
         PSLA \citep{gong2021psla} & IN+AS &  47.4 \\
         AST \citep{gong2021ast} & IN+AS &  48.5 \\
         HTS-AT \citep{chen2022hts} & IN+AS &  48.7 \\
         PaSST \citep{koutini2021efficient} & IN+AS &  49.6  \\
        \hline
         \our{} (5 models) & AS & 50.4  \\
         \our{} (10 models) & AS & \textbf{50.6}  \\
         \bottomrule
       
    \end{tabular}
    }
\end{table}

\subsection{Comparing with the SOTA Ensemble  Models}
Table~\ref{tab:ensemble} shows the comparison of the ensemble-model performance of our \our{} pre-trained models and the previous SOTA models on AS-2M.
We first ensemble all the five AS-2M fine-tuned \our{} models that are listed in Table~\ref{tab:tokenizer}, and denote it as \our{} (5 models). As shown in the table, without using any external supervised data (e.g. ImageNet), our \our{} (5 models) significantly outperforms the previous best ensemble models by 0.8 mAP. 
Then, we rerun the AS-2M fine-tuning of the five \our{} SSL models with a learning rate of 5e-5 for 100k training steps, and ensemble all the ten AS-2M fine-tuned models. The \our{} (10 models) can further improve the ensemble results and achieve 50.6 SOTA mAP performance.

\section{Conclusion}
In this paper, we propose \our{}, an iterative audio pre-training framework
a self-supervised model for audio representation learning. 
Different from the previous audio SSL models that employ reconstruction loss as the pre-training objective,
we present a self-distilled tokenizer to convert continuous audio signals into discrete labels, enabling the classic mask and discrete label prediction pre-training.
\our{} achieve superior performance across six audio and speech classification tasks and set new state-of-the-art results on AudioSet-2M and ESC-50 benchmarks.
Further analysis via visualization illustrates the pre-training targets of \our{} models are more robust to disturbances and aligned with the semantics than reconstruction-based audio SSL models, which indicates the effectiveness of the self-distilled tokenizer and accounts for the superiority of the proposed audio pre-training framework.

In the future, we would like to scale up the model size and pre-training data to further push the limits of audio classification. In addition, it is interesting to study the multi-modality field by combining audio with vision and language. 
\bibliography{neurips_2022}
\bibliographystyle{plainnat}

\appendix

\section{Appendix}

\subsection{Hyperparamter Settings}
\label{ssec:hyper}
Table~\ref{tbl:hyperparams} shows the detailed hyperparameters that are used for \our{} acoustic tokenizer training, audio SSL model pre-training and fine-tuning, which are adapted from the previous works \citep{xu2022masked,chen2022wavlm,beitv2}.

\begin{table}[ht]
\centering
\resizebox{1.0\columnwidth}{!}{
\begin{tabular}{l|c|c|c|c|c|c|c|c|c}
\toprule
 \multirow{2}{*}{Hyperparameters} & \multicolumn{2}{c|}{Tokenizer Training} & Model Pre-Training &  \multicolumn{6}{c}{Model Fine-Tuning} \\
 &  SSL Teacher &  SL Teacher &  AS-2M  &  AS-2M &  AS-20K &  ESC &  KS1 &  KS2 &  ER  \\
\midrule
Optimizer & \multicolumn{9}{c}{AdamW \citep{loshchilov2017decoupled}} \\
Optimizer Momentum & \multicolumn{9}{c}{$\beta_1=0.9, \beta_2=0.98$} \\
Weight decay & \multicolumn{9}{c}{0.01} \\
Learning Rate Schedule & \multicolumn{3}{c|}{Linear Decay} & \multicolumn{6}{c}{Cosine} \\
Steps & \multicolumn{3}{c|}{400K}  & 50K & \multicolumn{5}{c}{80K} \\
Warmup epochs & \multicolumn{3}{c|}{32K}  & 5K & \multicolumn{5}{c}{8K} \\
GPU & \multicolumn{2}{c|}{8} & 16  & 16 & \multicolumn{5}{c}{4} \\
Batch size (s) & \multicolumn{2}{c|}{1.4K} & 5.6K & 6.4K & 800 & 300 & \multicolumn{2}{c|}{100} & 300 \\
Layer-wise learning rate decay & \multicolumn{2}{c|}{1.0} & 1.0 & 0.6 & 0.3 & 0.2 & \multicolumn{2}{c|}{0.3}  & 1.0 \\
Peak learning rate & 5e-5 & 5e-4 & 5e-4  & 1e-4 & \multicolumn{2}{c|}{3e-5} & \multicolumn{2}{c|}{1e-4} & 3e-5  \\
\midrule
Weighted Sampling & \multicolumn{3}{c|}{\xmark} & \cmark  & \multicolumn{2}{c|}{\xmark} & \cmark*  & \multicolumn{2}{c}{\xmark} \\
Dropout \citep{srivastava2014dropout} & \multicolumn{3}{c|}{0.1} & \multicolumn{6}{c}{0.0}\\
Layer Dropout & \multicolumn{3}{c|}{0.0} & \multicolumn{6}{c}{0.1} \\
Roll Augmentation & \multicolumn{3}{c|}{\xmark} & \multicolumn{3}{c|}{\cmark} & \multicolumn{2}{c|}{\xmark} & \cmark \\
SpecAug \citep{park2019specaugment} & \multicolumn{3}{c|}{N/A} & \multicolumn{2}{c|}{0.3} & 0.2 & \multicolumn{2}{c|}{0.3} & 0.15 \\
Mixup \citep{zhang2017mixup} & \multicolumn{2}{c|}{N/A} & 0.0 & \multicolumn{2}{c|}{0.8} & 0.0 & \multicolumn{2}{c|}{0.8} & 0.0  \\
Multilabel & \multicolumn{2}{c|}{N/A} & \xmark & \multicolumn{2}{c|}{\cmark} & \xmark & \multicolumn{2}{c|}{\xmark} & \xmark \\
Loss Function & \multicolumn{2}{c|}{CosineSimilarity} & CE & \multicolumn{2}{c|}{BCE} & CE & \multicolumn{2}{c|}{BCE} & CE \\
\midrule
Dataset Mean for Normalization  & \multicolumn{5}{c|}{15.41663} & 11.72215 & 11.43905 & 11.41045 & 12.0889 \\
Dataset Std for Normalization & \multicolumn{5}{c|}{6.55582} & 10.60431 & 5.64913 & 5.67857 & 4.29147 \\
\bottomrule
\end{tabular}
}
\caption{
Hyperparameters of \our{} acoustic tokenizer training, audio SSL model pre-training and fine-tuning. 
*We balance each class to 50\% of the size of the unknown class for each training epoch.
}
\label{tbl:hyperparams}
\end{table}

\end{document}